\documentclass[a4paper,oneside,final,notitlepage,onecolumn]{article}
\usepackage{amsfonts}
\usepackage{epsf}
\usepackage{t1enc}
\usepackage[latin2]{inputenc}

\begin{document}

\newtheorem{theorem}{Theorem}[section]
\newtheorem{lemma}{Lemma}[section]
\newtheorem{proposition}{Proposition}[section]
\newtheorem{corollary}{Corollary}[section]
\newtheorem{conjecture}{Conjecture}[section]
\newtheorem{example}{Example}[section]
\newtheorem{definition}{Definition}[section]
\newtheorem{remark}{Remark}[section]
\newtheorem{exercise}{Exercise}[section]
\newtheorem{axiom}{Axiom}[section]
\newtheorem{condition}{Condition}[section]
\renewcommand{\theequation}{\thesection.\arabic{equation}} 

\author{\small Istv\'{a}n R\'{a}cz 
\\ 
\small MTA KFKI, Részecske- és Magfizikai Kutatóintézet\\
\small  H-1121 Budapest, Konkoly Thege Miklós út 29-33.\\ 
\small Hungary\\ 
\small  E-mail: istvan@sunserv.kfki.hu}

\date{\small \today}

\title{{\bf Symmetries of spacetime and their relation to initial value
problems}\thanks{%
This research was supported in part by OTKA grant T034337.} }

\maketitle

\begin{abstract}
We consider covariant metric theories of coupled gravity-matter
systems satisfying the following two conditions: First, it is
assumed that, by a hyperbolic reduction process, a system of first
order symmetric hyperbolic partial differential equations can be
deduced from the matter field equations. Second, gravity is supposed
to be coupled to the matter fields by requiring that the Ricci tensor
is a smooth function of the basic matter field variables and the
metric. It is shown then that the ``time'' evolution of these type of
gravity-matter systems preserves the symmetries of initial data
specifications.
\end{abstract}

\parskip 5pt

\section{Introduction}
\setcounter{equation}{0}

In a recent paper \cite{r} the necessary and sufficient conditions
ensuring the existence of Killing vector fields in covariant metric
theories of coupled gravity matter systems were investigated. In that
work matter fields were assumed to satisfy quasi-linear wave
equations and to possess minimal coupling to gravity. In addition,
gravity was supposed to be coupled to the matter fields by requiring
that the Ricci  tensor is a function of the basic matter field
variables, their first covariant derivatives and the metric. Within
this framework it was demonstrated that symmetries of initial data
specifications are preserved by ``time'' evolution.

The pragmatic aim in \cite{r} was to justify some of the claims of
\cite{frw,r1} concerning the existence of Killing vector fields in the
characteristic initial problem in case of Einstein--Klein-Gordon,
Einstein--[non-Abelian] Higgs or Einstein-[Maxwell]-Yang-Mills-dilaton
systems. The matter field variables\footnote{%
In case of gauge fields they are yielded by suitable gauge fixing (see
e.g. \cite{r,r1} for further details).} of these
particular Einstein-matter systems all satisfy quasi-linear wave
equations. However, the evolution of various matter fields, such as 
perfect- or dissipative fluids and spinor fields, is not
governed by quasi-linear wave equations; rather they satisfy first
order symmetric hyperbolic equations.

Clearly, it is of obvious physical interest to enlarge the framework
of \cite{r} to include matter fields satisfying equations of this more
general type and to determine those  conditions which guarantee the
existence of a Killing vector field for the associated  gravity-matter
systems. This is, in fact, our principal aim in this paper which is
organized as follows. The next section is to specify the class of
gravity-matter systems to which our main result applies. In
particular, a representation of the physical fields and the relevant
basic field equations are chosen so that well-posed initial value
problem can be associated with the selected gravity-matter
systems. Section \ref{kvf} is to study all the equations relevant for
the `evolution' of the Lie derivatives, with respect to a  `candidate
Killing vector field', of our basic variables. This section includes
our main result justifying the claim that the ``time'' evolution of
the selected gravity-matter systems preserves the symmetries of
initial data specifications. Finally, section \ref{conc} contains our
concluding remarks. In particular, issues associated with the
evolution of matter fields on fixed geometrical backgrounds and the
possible affection of a non-dynamical field on the evolution of
symmetries are considered.

\section{The gravity-matter systems}\label{spec}
\setcounter{equation}{0}

In order to have a suitable framework for the study of spacetime
symmetries, first of all we give a mathematically precise
specification of the gravity-matter theories to which our main result
will apply. Obviously, it is impossible to recall details of all the
particular systems which have ever been studied. Instead, a general,
and thereby flexible, enough framework is chosen within which common
features of a  reasonably large set of particular theories can be
investigated. The model specified  below is highly influenced by the
fundamental study  of Geroch \cite{g}, where a universal treatment of
physical fields and equations is presented.

\subsection{The choice of basic variables and field equations}

The physical fields are supposed to be represented by smooth tensor
fields  on a smooth four-dimensional paracompact orientable manifold
$M$. In particular, the ``gravitational field'' is assumed to be
represented by a smooth Lorentzian metric field $g_{ab}$ on $M$ and
all the gravitational effects are expected to be described in terms of
this metric and covariant quantities derivable from it. It is known
that other basic variables such as tetrads or spin frames could also
be used to represent gravity.  However, since our eventual aim in this
work is to study spacetime symmetries the use of $g_{ab}$ suits the 
most.\footnote{%
Unless otherwise stated we shall use the  notation and conventions of
\cite{wald}.}

In general, tensor fields representing the matter content of a theory
will be denoted by  $T_{_{(i)}}^{\mathcal{A}}$, where  $i$ in round
brackets is a name index taking values from the set
$\{1,2,...,N\}$, and the capital script letter ${\mathcal{A}}$  stands
for a composite (abstract) index. The constituents of ${\mathcal{A}}$,
i.e. the exact form ${T_{_{(i)}}^{a_1...a_{k_i}}}_{b_1...b_{l_i}}$ of
the $(k_i,l_i)$ type tensor field $T_{_{(i)}}^{\mathcal{A}}$ will be
spelled out explicitly in all the non-self-explaining situations. The
matter fields might also have gauge dependence but, even if they have,
it will be assumed that suitable gauge choices have been made so that
they can be considered as fields on the spacetime manifold. Thereby
the relevant gauge or internal space indices will be suppressed.

The evolution of the fields $T_{_{(i)}}^{\mathcal{A}}$ is supposed to
be given in terms of partial differential equations. In particular,
$T_{_{(i)}}^{\mathcal{A}}$ are assumed to solve systems of first order
quasi-linear partial differential equations of the form
\begin{equation}
\sum_{(j)}{K_{_{(i)(j)}}^{\mathcal{A}e}}{}_{\mathcal{B}} \nabla_e
T_{_{(j)}}^{\mathcal{B}}+L_{_{(i)}}^{\mathcal{A}}=0, \label{me}
\end{equation}
were $\nabla_a$ denotes the unique (torsion free) covariant derivative
operator associated with the spacetime metric $g_{ab}$.  Moreover, the
coefficients ${K_{_{(i)(j)}}^{\mathcal{A}e}}{}_{\mathcal{B}}$ and
$L^{\mathcal{A}}_{_{(i)}}$  in (\ref{me}) are expected to be smooth
fields on a tensor bundle, $\mathcal T$, built up from the  fields
$T_{_{(i)}}^{\mathcal{A}}$ and $g_{ab}$.

It is worth emphasizing that this framework 
covers matter systems like Klein-Gordon fields, Maxwell fields, perfect
fluids and  various other  type of matter fields, along with their
conventional couplings (see for more details Appendix A of \cite{g}).  In
fact, it seems to cover virtually every partial differential equation
in physics.\footnote{%
In principal, spinor fields could also be
present among the allowed matter  field variables. This is particularly so,
because the Dirac equations possess the form of (\ref{me}). Moreover,
it was demonstrated by Friedrich and Rendall \cite{fr} that symmetric
hyperbolic equations can be built up from the Einstein-Dirac system.
The main purpose of this work, however, is to study the existence of
symmetries which requires the use of Lie derivatives of all  the
fields with respect to essentially arbitrary vector fields on $M$. On
the other hand, it is known that  the Lie derivative of spinor fields
cannot be defined (at least not in the conventional way) with respect
to arbitrary vector fields but only to conformal Killing vector
fields. Before the procedure described below can be extended to
Einstein-Dirac systems as well one should clear up e.g. how to make a
sensible variation of a composite metric-spinor system. Obviously,
there is no problem of varying the spinor field while the metric is
kept fixed. However, a procedure of complete generality, such as
`changing the metric while keeping the spinor field fixed' does not
exist.} In particular, quasi-linear wave equations of the form
\begin{equation}
\nabla^f\nabla_f {T}_{_{(i)}}^{\mathcal{B}} +
F_{_{(i)}}^{\mathcal{B}}(g_{cd},{T}_{_{(j)}}^{\mathcal{C}},
\nabla_c{T}_{_{(j)}}^{\mathcal{C}})=0,  \label{ql}
\end{equation}
can immediately be put into the form of (\ref{me}), by following a
standard procedure (see e.g. pages 76-77 of \cite{fr}) which starts, by
adding the first derivatives ${T}_{_{(N+i)}}^{\mathcal{A}}{}_e=
\nabla_e {T}_{_{(i)}}^{\mathcal{A}}$ to the basic variables. 

Note also that, as indicated above, in case of matter fields, like
Yang-Mills or Higgs fields with  a gauge freedom, to get the desired
framework a gauge fixing is required to be made. For instance, in
case of a Yang-Mills field the use of the generalized Lorentz gauge
condition is needed to put the Yang-Mills equations into the form of a
quasi-linear wave equation (see for details e.g. \cite{fri2,r}), which
can then be recast into the form of (\ref{me}) by the above mentioned
procedure.

\smallskip
 
According to the above assumptions matter fields are assumed to be
coupled to gravity through the dependence of 
${K_{_{(i)(j)}}^{\mathcal{A}e}}{}_{\mathcal{B}}$ and
$L^{\mathcal{A}}_{_{(i)}}$ on 
the metric, as well as, via the presence of the metric compatible
covariant derivative operator $\nabla_e$ in (\ref{me}). Concerning the
coupling of the geometry to matter fields, we shall require that the
Ricci tensor $R_{ab}$ be a smooth function of the matter field
variables and the metric,
\begin{equation}
R_{ab}=R_{ab}(T_{_{(i)}}^{\mathcal{A}},g_{mn}).\label{g1}
\end{equation}
Note that this assumption immediately imposes  a restriction, in
addition to (\ref{me}), on the fields $T_{_{(i)}}^{\mathcal{A}}$ via
the twice contracted Bianchi identity.

It is straightforward to see that the conditions so far have been made
are satisfied by all the `customary' Einstein-matter systems. In
particular, the present framework immediately covers the setting
applied in \cite{r} since the matter field variables used here
includes the first covariant derivatives  of the basic matter field
variables applied there.  Moreover, the selected model is general
enough to host e.g. the `conformally equivalent representation' of
higher curvature theories associated with a gravitational Lagrangian
that is a polynomial of the Ricci scalar (for further details see
e.g. \cite{jk}).

In addition to the above assumptions we shall impose the following
requirement which plays a significant role in the derivation
of our main result and which is usually made implicitly in most of
the related discussions.

\begin{condition}\label{con0}
The fields ${K_{_{(i)(j)}}^{\mathcal{A}e}}{}_{\mathcal{B}}$ and
$L^{\mathcal{A}}_{_{(i)}}$, along with the Ricci tensor when it is
viewed as a functional of the basic variables as the expression
appearing on the r.h.s. of (\ref{g1}), are assumed to be smooth
functionals on the tensor bundle $\mathcal T$ depending exclusively on
$T_{_{(i)}}^{\mathcal{A}}$ and $g_{ab}$,
\begin{equation}
{K_{_{(i)(j)}}^{\mathcal{A}e}}{}_{\mathcal{B}}=
{K_{_{(i)(j)}}^{\mathcal{A}e}}{}_{\mathcal{B}}  (T_{_{(k)}}
^{\mathcal{C}},g_{mn})   \hskip .5cm {\rm and }
\hskip .5cm L^{\mathcal{A}}_{_{(i)}}=L^{\mathcal{A}}_{_{(i)}}
(T_{_{(j)}}^{\mathcal{C}},g_{mn}),
\end{equation}
without having any explicit dependence on the points of $M$. 
\end{condition}

This assumption expresses the idea that the distinction between two
points of a spacetime manifold $M$ is rooted in the difference between
the values of some physical field at that points. In other words, the
points of the spacetime $(M,g_{ab})$ are assumed to be identified by
observing some physical fields not by somehow ``perceiving the points
of $M$  themselves directly'' \cite{g}. Notice, however, that whenever
the dynamics of  certain fields $T_{_{(i)}}^{\mathcal{A}}$ on a fixed
background spacetime is considered  the coefficients
${K_{_{(i)(j)}}^{\mathcal{A}e}}{}_{\mathcal{B}}$ and
$L_{_{(i)}}^{\mathcal{A}}$ will immediately  pick up an `explicit'
spacetime dependence through that of $g_{ab}$.

Note finally that the relations (\ref{me}) and (\ref{g1}), along with
condition \ref{con0},  manifest an important property of the composite
gravity-matter system. Namely, they guarantee that the coupling of
all the subsystems rests upon only zero order terms of the
relevant basic variables.

\subsection{Hyperbolic reduction of the field equations}

A key feature of the field equations of physics is that a system of
partial differential equations can be deduced from them so that, for
suitable initial data specifications, this equations yield the `time' 
evolution of the relevant system. This subsection is to impose
further conditions which guarantee the well-posed initial value
problems can be associated with the selected gravity-matter theories.

\subsubsection{Symmetric hyperbolic systems for the matter fields}

To attain `existence and uniqueness results' relevant for an initial
value problem associated with the gravity-matter systems of the above
type first
the basic field equations have to be put into a specific form of PDEs
for which  existence and uniqueness of solutions is guaranteed. In
this work first order symmetric hyperbolic PDEs are used  for this
purpose which are specified as follows \cite{fr}:

Consider, as our unknown, a function $\Phi$ defined on an open subset of
$\mathbb{R}^4$ and assume that it takes values in a finite dimensional
vector space $V$.  The hyperbolic systems we are interested in now are
of the form
\begin{equation}\label{sh}
A^\gamma\partial_\gamma \Phi + B=0,
\end{equation}
where $A^\gamma=A^\gamma(x,\Phi)$ are supposed to be smooth functions
defined on an open subset of  $\mathbb{R}^4\times V$ taking values in
the vector space $L(V)$ of linear maps of $V$ to itself, while
$B=B(x,\Phi)$ is expected to be a smooth function on an  open subset of
$\mathbb{R}^4\times V$ with values in $V$. The basic requirement on
the system to be called {\it symmetric hyperbolic} is that the
$A^\gamma$ are symmetric, with respect to some inner product on $V$,
and there exists a 
covector field $\xi_\gamma$ such that $A^\gamma\xi_\gamma$ is positive
definite for all admissible values of $\Phi$.

\medskip

By imposing the following condition we shall restrict our
considerations to matter fields for which a symmetric hyperbolic
system can be deduced from (\ref{me}).

\begin{condition}\label{con}
Equation (\ref{me}) can be split into a symmetric hyperbolic system of
the form (\ref{sh}), in terms of a suitable variable $\Phi$ labeling
cross-sections of the  tensor bundle built up from the fields
$T_{_{(i)}}^{\mathcal{A}}$, and constraint equations which, along with
the possible gauge conditions, propagate under the evolution of the
associated hyperbolic system.
\end{condition}

Note that there is no known a general practical procedure which would
immediately  produce a symmetric hyperbolic system of the form
(\ref{sh}) from (\ref{me}). In fact, the actual way of deducing a
symmetric hyperbolic system (\ref{sh}) from (\ref{me}) and the
justification that the constraints also  propagate for each particular
case might be very complicated and most of the time requires very
special considerations. However, the actual way of getting the required
hyperbolic reduction of the selected matter system will not play any
role in the basic argument concerning the evolution of symmetries of
initial data specifications.

Notice also that even after imposing condition \ref{con} the framework
we have still covers matter systems like Klein-Gordon, dilaton, Higgs,
Maxwell-Yang-Mills fields fields and various fluids along with the
conventional couplings of these systems. In fact, this setting is
suitable to host virtually all the physically relevant matter systems
for which a well-posed initial value problem can ever be formulated.

\subsubsection{Hyperbolization of the gravitational part}

To get symmetric hyperbolic equations relevant for the selected
gravity-matter  system we need to take care beside the hyperbolic
reduction of (\ref{me}) that of (\ref{g1}), as well. It is known
\cite{g,fr} that, due  to the general covariance of the considered
metric theories of gravity, there is no way to get hyperbolic
evolution from the coupled field equations (\ref{me}) and (\ref{g1})
without fixing the gauge associated with the diffeomorphism
invariance. One of the possibilities\footnote{%
For an excellent review of all the related issues see \cite{f1}.} to
make such a gauge fixing rests on the form of the Ricci tensor 
$R_{ab}$ given in local coordinates $x^\alpha$ as 
\begin{equation}\label{rilo}
R_{\alpha\beta}=-\frac{1}{2}g^{\mu\nu}\partial_\mu\partial_\nu
g_{\alpha\beta}+ g_{\delta(\alpha}\partial_{\beta)}\Gamma^\delta+
H_{\alpha\beta}\left(g_{\varepsilon\rho}, \partial_{\gamma}
g_{\varepsilon\rho}\right),
\end{equation}
and it is done by simply replacing $\Gamma^\delta
=g^{\gamma\rho}  {\Gamma^\delta}_{\gamma\rho}$ in (\ref{rilo}) with 
arbitrary `gauge source functions' $f^\delta$ (chosen so that
$f^\delta=\Gamma^\delta$ is satisfied on initial hypersurfaces). The
equations yielded by substituting, in local coordinates, the r.h.s. of
(\ref{rilo}) for the l.h.s of (\ref{g1}), with $\Gamma^\delta$
replaced by $f^\delta$, are refereed to as {\it reduced gravity
equations}. These equations can be put into the form of a first order
symmetric hyperbolic equation by introducing all first order
derivatives of $g_{\alpha\beta}$ 
as new variables and let $\Phi$ consists of $g_{\alpha\beta}$ together
with these derivatives. Then by making use of the reduced gravity equations a
system of symmetric hyperbolic equations for $\Phi$ can be built up by a
standard procedure (see for more details e.g. \cite{ch,fr}).

\subsubsection{Hyperbolic reduction of the coupled system}

Upon having a hyperbolic reduction of the matter and gravity equations
separately, it is important to know whether the coupled 
equations also form a symmetric hyperbolic system. The
answer is in the affirmative. This can be justified by recalling that
the coupling of the matter field equations (\ref{me}) -- which by {\sl
condition \ref{con}} can be cast into a symmetric hyperbolic system of
the form (\ref{sh}) -- and that of a fixed
hyperbolic reduction of (\ref{g1}) -- written in first order symmetric
hyperbolic form -- is only by terms of order zero, hence, the
combined system is also symmetric hyperbolic. Therefore, local
existence and uniqueness results immediately apply to the relevant
reduced gravity-matter equations in case of regular initial
value problems.

There is only one significant requirement the applied initial value
problem  has to satisfy:  The existence and uniqueness of solutions to
symmetric hyperbolic equations  is expected to be guaranteed within
its framework in the smooth setting.  The initial value problems which
satisfy this condition are referred to as {\it regular initial
value problems}. Immediate examples for appropriate  initial value
problems are the standard Cauchy problem (see e.g. Ref. \cite{ch}) and
also the  characteristic initial value problem associated with an
initial hypersurface  represented by either the union of two smooth
null hypersurfaces intersecting on a 2-dimensional spacelike surface
\cite{mzh,rendall} or a characteristic cone
\cite{fried,cag,rendall}. Since no further requirement on the initial
value problem  is used anywhere in the derivation of our results we
shall not make a definite choice  among these regular initial
value problems. Thereby the following  notation will be  applied on
equal footing to any of these initial value problems: The initial
hypersurface will be denoted by $\Sigma$, while the initial data on
$\Sigma$ will be represented  by the basic variables in square
brackets. In particular,  an initial data specification for an
equation of the form (\ref{sh}) is represented by the pair
$(\Sigma,[\Phi])$. 

As mentioned above the hyperbolic reduction is done by splitting the
basic equations to pure evolution equations and constraint
ones. In fact, the constraint equations, along with the gauge
conditions, manifest certain (differential) identities that have to be
satisfied by $T_{_{(i)}}^{\mathcal{A}}$ and $g_{ab}$. In particular,
they impose restrictions on the possible initial data sets. Thereby,
having a hyperbolic reduction of the coupled system suitable initial data
consistent with the relevant gauge conditions and satisfying all the
constraints has to be arranged.  Finally, it has also to be
demonstrated that the gauge
conditions and the constraints propagate with the `time' evolution of  the
relevant coupled reduced system. In the present case, by making use of
a suitable combination of {\sl condition \ref{con}} and the twice contracted
Bianchi identity (see also page 20 and pages 76-77 of \cite{fr}), it can be
justified that the gauge conditions and the constraints will be
satisfied throughout the domain of dependence associated with a
solution to the coupled evolution 
equations if they hold on the initial data surface.

\section{The evolution of symmetric initial data specifications}\label{kvf} 
\setcounter{equation}{0}

Consider a gravity-matter system of the type specified in the previous
section satisfying {\sl conditions} \ref{con0} and \ref{con}. Assume
that a regular initial value problem has been chosen, moreover, a
suitable initial data set relevant for the coupled symmetric
hyperbolic evolution equations also satisfying the pertinent gauge
conditions and constraint equations has also been selected. Recall
that then we immediately have a detailed knowledge of the metric  and
matter fields on the initial data surface $\Sigma$ since in any of the
regular initial value problems  $T_{_{(i)}}^{\mathcal{A}}$ and
$g_{ab}$, along with their derivatives up to arbitrary order, can
always be determined on $\Sigma$ by making use of an initial data
specification, $(\Sigma,[T_{_{(i)}}^{\mathcal{A}}],[g_{ab}])$, and the
field equations.  Clearly, it is of obvious interest to identify those
initial data sets that are guaranteed to evolve to configurations
possessing certain type of symmetries.  The rest of this paper is to
make it clear under what conditions the symmetries, and more
importantly what sort of symmetries, of initial data sets will be
inherited by the corresponding solutions.

There are various possibilities which make it sensible to consider an
initial data specification to be symmetric. The most straightforward
one is the following: An initial data  set
$(\Sigma,[T_{_{(i)}}^{\mathcal{A}}],[g_{ab}])$ is considered to be
symmetric if there exists a diffeomorphism $\psi: \Sigma \rightarrow
\Sigma$ of the initial hypersurface $\Sigma$ onto itself  so that it
leaves the initial data invariant. These type of symmetries of initial
data specifications were considered in \cite{fr}, for the case of
Einstein-vacuum systems. Based essentially on the uniqueness of
maximal Cauchy developments, it was demonstrated there that such a
symmetry of a vacuum initial data set will be inherited by the
corresponding solution. By making use of a straightforward adaptation
of this argument (see, for more details, pages 90-91 of \cite{fr}) to
the case of 
gravity-matter systems investigated here, it can also be justified
that a symmetry, of the considered type, of an initial data set gives
also rise to a symmetry of the associated Cauchy development,
i.e. there will exist a diffeomorphism $\Psi: M \rightarrow M$ with
$\Psi_*  T_{_{(i)}} ^{\mathcal{A}} = T_{_{(i)}}^{\mathcal{A}}$ and
$\Psi_*g_{ab} =g_{ab}$ with $\Psi\vert_\Sigma=\psi$.

Since there are symmetries of Cauchy developments which does not
necessarily map  any initial data surfaces onto themselves it is of
obvious interest to know whether the traces of these more general type
of symmetries can also be recognized by making use of initial data
specifications. It seems to be hopeless to find an evidence for the
existence of these more general type of  symmetries in the case of
discrete symmetry transformations. However, whenever a gravity-matter
system admits a  Killing vector field, i.e. there exists, at least
locally, a one-parameter family of isometry actions in a
neighbourhood of $\Sigma$, the procedure described in detail in the
following  section can be applied.

Before approaching to this point there are two comments in order.
Firstly, I would like to emphasize the complementary character of the 
procedures associated with the above mentioned two types of symmetries.
The results relevant for the more general type of
symmetries obviously cover the special case whenever an initial data
specification admits a one-parameter family of diffeomorphisms
$\psi_\tau: \Sigma \rightarrow \Sigma$ leaving the initial data
invariant.  However, it has its drawback not being able to handle the
case of discrete isometry actions which, on the other hand, can be
covered by the indicated generalization of the method of Friedrich and
Rendall \cite{fr} in the particular case when the initial data surface
is mapped onto itself by the associated diffeomorphism. 

Secondly, the results presented below are due to straightforward
adaptations  of earlier  methods applied in the Einstein-vacuum case
to study the linearization stability problem by Moncrief
\cite{vm1,vm2} (see also \cite{bc,chr}).  Note that these results
provide an immediate generalization of the main assertion of \cite{r},
and motivates a possible further strengthening of the conclusion of
\cite{k}, since the results covered by \cite{r,k} are relevant merely
for gravity-matter systems  with matter fields satisfying quasi-linear
wave equations.

\subsection{The construction of a candidate Killing vector field}

It is known that if we had a Killing vector field $K^a$, satisfying
the Killing equation,
\begin{equation}
\mathcal{L}_K g_{ab}=\nabla_a K_b+\nabla_b K_a= 0
\end{equation}
and the integrability condition,
\begin{equation}
\nabla _a\nabla _bK_c+{R_{bca}}^dK_d=0,  \label{RK}
\end{equation}
of the 2-form field $\Xi_{ab}=\nabla_{a} K_{b}$
this Killing vector field would be completely determined by the values
of $K_a$ and  $\nabla_a K_b$ at any point of $M$ since the above two
equations imply a system of ODE's for the components of $K_a$ and
$\nabla_a K_b$ along any $C^1$ curve. However, the existence of a
Killing vector field cannot  be proven in this way. Nevertheless, the
contraction, 
\begin{equation}
\nabla ^e\nabla _eK^a+{R^a}_dK^d=0,  \label{LK}
\end{equation}
of (\ref{RK}), which is a linear homogeneous wave  equation for $K^a$,
provides the means to construct a `candidate' Killing vector
field. Obviously, any Killing vector field satisfies (\ref{LK})  but
not all of its solutions will give rise to a Killing vector field.

\subsection{Evolution of the Lie derivatives of the basic
variables}

The conditions ensuring the existence of initial data $[K^a]$ on an
initial hypersurface $\Sigma$ so that the unique solution $K^a$ of
(\ref{LK}) will be a Killing vector field are necessarily given in terms of
restrictions on the values of the fields $T_{_{(i)}}^{\mathcal{A}}$
and $g_{ab}$ on $\Sigma$. The relevant requirements can be read off the
evolution equations for the Lie derivatives of our basic variables. The
remaining part of this section  is to identify these conditions and
thereby we shall prove the following:

\noindent{\bf Theorem:} {\ }{\it
Let $(M,g_{ab})$ be a spacetime associated with a gravity-matter
system as it was specified in section \ref{spec}. Denote by
$D[\Sigma]$ the domain of dependence of an initial  hypersurface
$\Sigma$ associated with a regular initial value problem. Then there
exists a non-trivial Killing vector field $K^a$ on $D[\Sigma]$, so
that the matter fields are also invariant, i.e.
$\mathcal{L}_KT_{_{(i)}}^{\mathcal{A}}=0$, if there exists a
non-trivial initial data set $[K^a]$ for (\ref{LK})  so that
$\mathcal{L}_{K}T_{_{(i)}}^{\mathcal{A}}$, $\mathcal{L}_{K}{g}_{ab}$
and $\nabla_c( \mathcal{L}_{K}{g}_{ab})$ all vanish identically on
$\Sigma$.}

\medskip

Before presenting the proof of this theorem let us emphasize certain
points relevant for the applied setting.  First of all, the practical
problem of finding initial data $[K^a]$ for (\ref{LK}) -- which in
each particular case is the hard part of the work -- is not considered
here. It is, however, important to note that, upon having any sort of
initial data $[K^a]$ for (\ref{LK}), the validity of our conditions,
i.e. the vanishing of $\mathcal{L}_{K}T_{_{(i)}}^{\mathcal{A}}$,
$\mathcal{L}_{K}{g}_{ab}$ and $\nabla_c( \mathcal{L}_{K}{g}_{ab})$,
can be immediately justified. This is a consequence of the fact that
knowledge of $[K^a]$, along with the equation (\ref{LK}), suffices to
determine derivatives of $K^a$ up to any order on $\Sigma$.  As
already mentioned above, the fields  $T_{_{(i)}}^{\mathcal{A}}$ and
$g_{ab}$, along with their derivatives up to arbitrary order, can also be
determined on the initial hypersurface $\Sigma$ in any regular initial
value problem. Thereby the existence of a local
one-parameter group of diffeomorphisms acting on the Cauchy
development which leaves the matter fields and the geometry invariant
can  be read off the initial data for $T_{_{(i)}}^{\mathcal{A}}$ and
${g}_{ab}$ in advance of having the relevant solution explicitly.

Secondly, in case of gauge fields it might seem to be too strong to require
the vanishing of the Lie derivative of the field variable. It is
known, for instance, that in the particular case of a Yang-Mills field
the invariance of a vector potential $A_a$ (taking values in a Lie
algebra $\mathfrak{g}$) under the action of a 
one-parameter group of diffeomorphisms associated with a vector field
$K^a$ implies only that $\mathcal{L}_{K}A_{a}$ is equal to the gauge
covariant derivative $D_a$ of a $\mathfrak{g}$-valued function $W$,
i.e. $\mathcal{L}_{K}A_{a}=D_aW$. This general invariance property
plays an important role whenever there are several symmetries acting
simultaneously on a given spacetime. Note, however, that in the
present situation -- when we look for a single candidate Killing vector
field determined by (\ref{LK}) and suitable initial data specifications
-- the function $W$ can be gauged away (see, e.g., page 19 of \cite{fm}) so
that the invariance of an adapted vector potential $A_a$ can, in fact,
without loss of generality, be expressed as $\mathcal{L}_{K}A_{a}=0$. 

\noindent
{{\bf Proof:}}{\ } To start off consider a  vector field $K^a$
satisfying (\ref{LK}) but which is kept otherwise to be
arbitrary. Then, by taking 
the covariant derivative of (\ref{LK}), commuting derivatives and
applying the contracted  Bianchi identity, it can be shown that
$\mathcal{L}_{K}{g}_{ab}$ satisfies the equation
\begin{equation}
\nabla ^e\nabla _e\left(\mathcal{L}_Kg_{ab}\right)=
-2\mathcal{L}_KR_{ab}+2{{R^e}_{ab}}^{f}(\mathcal{L}_Kg_{ef})
+2R_{\;(a}^e(\mathcal{L}_Kg_{b)e}). \label{wsab}
\end{equation}
Moreover, by taking the Lie derivative of (\ref{g1}) we get\footnote{%
Whenever $T^{\mathcal{A}}$ and $S^{\mathcal{B}}$ are tensor  fields of
type $(k,l)$ and $(m,n)$, respectively, $(\partial
T^{\mathcal{A}}/\partial S^{\mathcal{B}})$ is considered  to be a
tensor field of type $(k+n,l+m)$. Accordingly, the contraction
$(\partial T^{\mathcal{A}}/\partial S^{\mathcal{B}})\mathcal{L}_K
S^{\mathcal{B}}$  is again a tensor field of type $(k,l)$.}
\begin{equation}
\mathcal{L}_KR_{ab}=\sum_{(i)}\left(\frac{\partial R_{ab}}{\partial
T_{_{(i)}}^{\mathcal{A}}}\right) \mathcal{L}_KT_{_{(i)}}
^{\mathcal{A}} +\left( \frac{\partial R_{ab}} {\partial g_{ef}}\right)
\mathcal{L}_Kg_{ef}. \label{lie1}
\end{equation}
Then, in virtue of (\ref{wsab}) and (\ref{lie1}),
$\mathcal{L}_Kg_{ab}$ satisfies an equation of the form
\begin{equation}
\nabla ^e\nabla _e\left(\mathcal{L}_Kg_{ab}\right)=
{P}_{ab}(\mathcal{L}_Kg_{cd})+\sum_{(i)}{{Q}_{_{(i)}}}
{}_{ab}(\mathcal{L}_KT_{_{(i)}}^{\mathcal{A}})
\label{evol1}   
\end{equation}
where ${P}_{ab}$ and ${{Q}_{_{(i)}}}{}_{ab}$ are linear
and homogeneous functions of their indicated arguments.

\smallskip

The next step is to show that $\mathcal{L}_KT_{_{(i)}}^{\mathcal{A}}$
satisfy an equation analogous to (\ref{me}) so that the coefficient of
the principal part of the relevant equation is equal to that of
(\ref{me}). To achieve this take the Lie derivative of (\ref{me}) with
respect to the vector field $K^a$. This yields the relation
\begin{eqnarray}
\sum_{(j)} {K_{_{(i)(j)}}^{\mathcal{A}e}}{}_{\mathcal{B}}\left[
\mathcal{L}_K\left(\nabla_e T_{_{(j)}}^{\mathcal{B}}\right)\right] &+&
\sum_{(j)} \left\{\nabla_e   T_{_{(j)}}^{\mathcal{B}}\left[ \sum_{(k)}
\left(\frac{\partial
{K_{_{(i)(j)}}^{\mathcal{A}e}}{}_{\mathcal{B}}}{\partial
T_{_{(k)}}^{\mathcal{C}}}\right) \mathcal{L}_KT_{_{(k)}}
^{\mathcal{C}} + \left(\frac{\partial {K_{_{(i)(j)}}
^{\mathcal{A}e}}{}_{\mathcal{B}}} {\partial g_{cd}}\right)
\mathcal{L}_K g_{cd} \right]\right\} \nonumber \\ &+&\sum_{(l)}
\left(\frac{\partial L_{_{(i)}}^{\mathcal{A}}}{\partial
T_{_{(l)}}^{\mathcal{C}}}\right) \mathcal{L}_KT_{_{(l)}}^{\mathcal{C}}
+ \left(\frac{\partial L_{_{(i)}}^{\mathcal{A}}}{\partial  g_{cd}}\right)
\mathcal{L}_K g_{cd}  =0. \label{lie}
\end{eqnarray}
To put (\ref{lie}) into the form of (\ref{me}) the commutation
relation of the operators $\mathcal{L}_K$ and  $\nabla_a$ is needed to
be used which reads as
\begin{equation}
\mathcal{L}_K\left(\nabla_cT_{_{(i)}}^{\mathcal{A}}\right)=
\nabla_c\left(\mathcal{L}_KT_{_{(i)}}^{\mathcal{A}}\right) +
\sum_{s=1}^{k_i} \left(T_{_{(i)}}^{\mathcal{A}}\right){}_{[a_s]}^e
\left[\nabla\mathcal{L}_Kg\right]{{{}_e^{a_s}}}_{c}  -\sum_{t=1}^{l_i}
\left(T_{_{(i)}}^{\mathcal{A}}\right){}^{[b_t]}_e
\left[\nabla\mathcal{L}_Kg\right]{{{}_{b_t}}^e}_{c},\label{com}
\end{equation}
where $\left(T_{_{(i)}}^{\mathcal{A}}\right){}_{[a_s]}^e$ and
$\left(T_{_{(i)}}^{\mathcal{A}}\right){}^{[b_t]}_e$ stand for
$T_{_{(i)}}{}^{a_1...e...a_{k_i}}_{\hskip .45cm {s \atop \smile}
\hskip .55cmb_1...b_{l_i}}$ and  $T_{_{(i)}}{}^{a_1...a_{k_i}\hskip
.35cm {t \atop \smile} \hskip .6cm}_{\hskip .9cmb_1...e...b_{l_i}}$,
respectively; moreover, the notation
\begin{equation}
\left[\nabla\mathcal{L}_Kg\right]{{{}_{a}}^c}_{b}=\frac{1}{2}g^{cf}
\left\{\nabla_{a}\left(\mathcal{L}_Kg_{fb}\right)+
\nabla_b\left(\mathcal{L}_Kg_{{a}f}\right)
-\nabla_f\left(\mathcal{L}_Kg_{{a}b}\right)\right\}
\end{equation}
has been applied. Then, in virtue of (\ref{lie}) and
(\ref{com}),  the desired equation reads as
\begin{equation}
\sum_{(j)}{K_{_{(i)(j)}}^{\mathcal{A}e}}{}_{\mathcal{B}}
\nabla_e\left(\mathcal{L}_K T_{_{(j)}}^{\mathcal{B}} \right)
+{\widehat L}^{\mathcal{A}}_{_{(i)}}=0, \label{evol2}
\end{equation}
where the coefficients of the terms $\nabla_e(\mathcal{L}_K
T_{_{(j)}}^{\mathcal{B}})$ are exactly that of $\nabla_e
T_{_{(j)}}^{\mathcal{B}}$ in (\ref{me}), and, the ${\widehat
L}^{\mathcal{A}}_{_{(i)}}$ stand for a sum of terms which are linear
and homogeneous expressions of 
either of the variables $\mathcal{L}_K T_{_{(i)}}^{\mathcal{A}}$,
$\mathcal{L}_Kg_{ab}$ or $\nabla_c(\mathcal{L}_Kg_{ab})$,
respectively.

\smallskip

To complete our argument notice first that, in virtue of the
coincidence of the principal parts of (\ref{evol2}) and (\ref{me}),
along with condition \ref{con}, the same hyperbolization procedure
that was supposed to exist in case of  (\ref{me}) can also be used to
deduce from (\ref{evol2}) a symmetric hyperbolic system for the
variables $\mathcal{L}_K T_{_{(i)}}^{\mathcal{A}}$.\footnote{%
Recall
that a hyperbolic reduction of an equation of the type (2.1), whenever
it can be made, is achieved by a regular algebraic procedure (see
e.g. page 9 of \cite{g}) which depends on the principal part of the
equation exclusively. Thereby, however different the  source terms ${
L}^{\mathcal{A}}_{_{(i)}}$ and ${\widehat  L}^{\mathcal{A}}_{_{(i)}}$
might be, their difference has no influence on the hyperbolization of
equation (\ref{evol2}).}    Similarly, the linear wave equation
(\ref{evol1}) can be put into the form of a symmetric hyperbolic
system for the  variables $\mathcal{L}_Kg_{ab}$ and
$\nabla_c\mathcal{L}_Kg_{ab}$. Since the coupling of these two systems
happens only through terms of zero order the coupled system
immediately gives rise to a symmetric hyperbolic system for the fields
$\mathcal{L}_K T_{_{(i)}}^{\mathcal{A}}$, $\mathcal{L}_Kg_{ab}$ and
$\nabla_c(\mathcal{L}_Kg_{ab})$. In addition the `source term' of the
resulting system is a sum of expressions which are  linear and
homogeneous in either of these variables.  This property guarantees
that the combined equation necessarily have the identically zero
solution for vanishing initial data. Since the symmetric first order
system derived from (\ref{evol1}) and (\ref{evol2}) is linear and
homogeneous in the basic variables the constraints, that there might
be yielded by the relevant hyperbolic reduction of these equations,
are immediately satisfied by the identically zero solutions of the
coupled evolution equations. Hence whenever initial data $[K^a]$ can
be chosen on an initial hypersurface $\Sigma$ so that $\mathcal{L}_K
T_{_{(i)}}^{\mathcal{A}}$, $\mathcal{L}_Kg_{ab}$ and
$\nabla_c(\mathcal{L}_Kg_{ab})$ are zero on $\Sigma$ then each of the
fields $\mathcal{L}_K T_{_{(i)}}^{\mathcal{A}}$ and
$\mathcal{L}_Kg_{ab}$ will identically vanish throughout the domain
where the associated unique solution of (\ref{LK}) does exist.  To
see, finally, that this domain has to coincide with $D[\Sigma]$ note
that, since (\ref{LK})  is a linear homogeneous wave equation any
solution of it can be shown -- by making use of the `patching together
local solutions' techniques described e.g. on page 266 of \cite{mzhs}
-- to extend over the entire domain of dependence $D[\Sigma]$ of
$\Sigma$. \hfill\fbox{}

\section{Concluding remarks}\label{conc}
\setcounter{equation}{0} 
In virtue of the above result one may conclude that for any
gravity-matter systems 
satisfying conditions \ref{con0} and \ref{con} Killing symmetries of
initial data specifications $(\Sigma,[T_{_{(i)}}^{\mathcal{A}}],
[g_{ab}])$ are preserved by the ``time'' evolution. This section is to
consider situations where either of the conditions of the above
theorem fails to be satisfied. 

Let us start by considering the evolution of certain fields on a fixed
non-dynamical spacetime $(M,g_{ab})$.  Suppose that $K^a$ is a vector
field on $M$ satisfying (\ref{LK}) but the Lie derivative of  the
metric with respect to $K^a$ should not necessarily vanish. It is
straightforward to see that whenever $\mathcal{L}_Kg_{ab}$ does not
vanish the relevant form of  (\ref{evol1}) gives rise to a  constraint
equation which has to be satisfied by the Lie derivatives of fields
$T_{_{(i)}}^{\mathcal{A}}$. Furthermore, (\ref{evol2}) determining the
evolution of  $\mathcal{L}_KT_{_{(i)}}^{\mathcal{A}}$ will have
non-trivial source terms in consequence of the presence of the
non-vanishing fields $\mathcal{L}_Kg_{cd}$ and
$\nabla_e\mathcal{L}_Kg_{cd}$. This implies then  that despite of the
fact that the fields $T_{_{(i)}}^{\mathcal{A}}$ might be arranged to
possess certain type of symmetries on the initial hypersurface the
evolution of the system will suppress this property unless $K^a$ is a
Killing vector field of the fixed background spacetime.

The situation is very similar, although the equations simplify
considerably,  whenever $K^a$ is chosen to be a Killing vector field
for the fixed background geometry but the matter fields are kept to
be general. Then the non-invariance of the matter fields
-- represented by the Lie derivatives $\mathcal{L}_K
T_{_{(i)}}^{\mathcal{A}}$ --  `evolve' according to a linear
homogeneous equation of the form
\begin{equation}
\sum_{(j)}{K_{_{(i)(j)}}^{\mathcal{A}e}}{}_{\mathcal{B}}
\nabla_e\left(\mathcal{L}_K T_{_{(j)}}^{\mathcal{B}} \right)
+\sum_{(k)}{L_{_{(k)}}^{\mathcal{A}}}{}_{\mathcal{B}}
\left(\mathcal{L}_K T_{_{(k)}}^{\mathcal{B}} \right)=0, \label{evol22}
\end{equation}
where ${L_{_{(k)}}^{\mathcal{A}}}{}_{\mathcal{B}}$ are some smooth
fields, moreover, the Lie derivatives $\mathcal{L}_K
T_{_{(i)}}^{\mathcal{A}}$ have also to satisfy the additional constraint
\begin{equation}
\sum_{(i)}\left(\frac{\partial R_{ab}}{\partial
T_{_{(i)}}^{\mathcal{A}}}\right) \mathcal{L}_KT_{_{(i)}}
^{\mathcal{A}}=0.\label{evol222}
\end{equation}

\smallskip

An analogous argument applies whenever the geometry is chosen to be
dynamical and initial data is arranged for (\ref{LK}) so that
$\mathcal{L}_Kg_{cd}$ and $\nabla_e\mathcal{L}_Kg_{cd}$ vanish
identically on $\Sigma$, however, either of the Lie derivatives
$\mathcal{L}_K T_{_{(i)}}^{\mathcal{A}}$  is not zero there. Then the
unique solution $K^a$ of (\ref{LK}) will not give rise to a Killing
vector field of the coupled gravity-matter system and, in general,
neither of the fields $\mathcal{L}_K T_{_{(i)}}^{\mathcal{A}}$,
$\mathcal{L}_Kg_{cd}$ and $\nabla_e\mathcal{L}_Kg_{cd}$ will vanish
on succeeding Cauchy surfaces.

\smallskip 

Consider now a situation whenever condition \ref{con0} is not
satisfied, i.e. suppose that either of the fields ${K_{_{(i)}}^
{\mathcal{A}e}}{}_{\mathcal{B}}$,  $L^{\mathcal{A}}$ used in
(\ref{me}) or the expression appearing on the r.h.s. of (\ref{g1})
depends explicitly on the points of the spacetime. This can occur, for
instance, 
whenever there is an additional field present not taking part in the
evolution. It is straightforward to see that then the Lie derivative
of either of the fields ${K_{_{(i)}}^{\mathcal{A}e}}{}_{\mathcal{B}}$,
$L^{\mathcal{A}}$  or $R_{ab}$ will not be a homogeneous
expression of the Lie derivatives of the dynamical fields
exclusively. Instead, terms containing the Lie derivative of the
non-dynamical field appear in equations (\ref{evol1}) and
(\ref{evol2}).  Hence, unless this additional non-dynamical field is
itself invariant under the action of the one-parameter families of
diffeomorphisms associated with our to be Killing field $K^a$, these
terms behave as source terms yielding the loss of any
possible Killing symmetry of an initial data specification.

It is important to emphasize that the presented results also provide
means to justify or dispel certain assumptions (usually made
implicitly) which are applied e.g. in searching for exact or
numerical solutions of the type of gravity-matter systems investigated
here. For instance, if a system does not satisfy condition \ref{con0}
it straightforwardly follows from the above considerations that the
application of a dimensional reduction, based on the assumption that
the spacetime admits a global symmetry, might not be completely
consistent with the system under investigation. Clearly, it may happen
that  a gravity-matter system of this type has no solution which would
possess a (local) one-parameter group of isometry actions which also
would leave the matter fields to be invariant.

Let us finally mention that on a similar base the results presented in
this paper might be useful in testing the trustworthiness of a code in
numerical simulation of gravity-matter systems. In particular, in case
of theories satisfying conditions \ref{con0} and \ref{con} one can be
sure that something has to be wrong with a code if it does not
preserve (at least approximately) symmetries of initial data
specifications.

\vfill\eject

\end{document}